\documentclass[referee]{aa}
\usepackage{natbib,txfonts,graphicx}
\begin{document}
\title{Assessing the tilt of the solar magnetic field axis \\ through Faraday rotation observations}
\author{S. Mancuso \inst{1} \and  M.V. Garzelli \inst{2}}
\institute{INAF, Osservatorio Astronomico di Torino, via Osservatorio 20, 10025 Pino Torinese (To), Italy
\and Dipartimento di Fisica, Universit\`a di Milano and INFN, Sezione di Milano, via Celoria 16, 20133 Milano, Italy}
\mail{S. Mancuso: mancuso@oato.inaf.it ,  
M.V. Garzelli: garzelli@mi.infn.it .} 
\date{Received / Accepted}
 
\abstract
   {Faraday rotation measurements of extragalactic radio sources during coronal occultation allow assessment of both the electron density distribution and the three-dimensional magnetic field topology in the outer solar corona.}
   {We simulate the three-dimensional structure of both the coronal magnetic field and the electron density distribution in order to reproduce the excess Faraday rotation measures (RMs) of the occulted radio sources observed during solar activity minimum. In particular, we infer the tilt of the solar magnetic axis with respect to the rotation axis.}
   {We compare the output of the model with Very Large Array (VLA) radio polarimetric measurements of a sample of extragalactic sources observed in May 1997. Information on the magnetic field geometry can be retrieved by fine-tuning the set of model free parameters that best describe the observations.}
   {We find that predicted and observed Faraday rotation measures are in excellent agreement, thus supporting the model. Our best-fitting model yields a tilt angle $\theta_{RB}=3.3^{\circ}$ of the solar magnetic axis with respect to the solar rotation axis around Carrington Rotation 1923. This result is consistent with analogous but independent estimates computed from the expansion coefficients of the photospheric field observed at the Wilcox Solar Observatory (WSO).}
   
\titlerunning{Assessing the tilt of the solar magnetic field}
\maketitle

\section{Introduction}
Magnetic fields play a crucial role in structuring the solar corona and provide a key to the physical processes that drive coronal heating and activity. However, notwithstanding the progress in observational capabilities attained in recent years a direct measurement of the coronal magnetic field remains a very difficult and challenging problem.  In the past, various techniques have been applied to achieve this aim. The microwave gyroresonance magnetometry technique is used to measure strong (more than a few hundred Gauss) active region field strengths in the low corona (Gary \& Hurford 1994; Lee et al. 1999). Circular po\-la\-ri\-za\-tion radio measurements of bremsstrahlung emission allow the detection of the longitudinal component of even relatively weak magnetic fields (Zhang et al. 2002). 
At optical, infra-red and EUV wavelengths, direct, though line-of-sight (LOS) integrated information about the coronal magnetic field can be obtained from the Zeeman and the Hanle effects 
applied to corona\-gra\-phi\-cally observed lines of highly ionized species (Lin et al. 2000, 2004; Raouafi et al. 2002). 

Radio propagation techniques are also important in ob\-tai\-ning indirect information about the properties of the solar coronal plasma (e.g., Mancuso et al. 2003).
So far,  
at least 
in the outer corona, the best remote-sensing technique currently available exploits the Faraday effect, which is a well-understood physical process. As radio waves travel through a magnetized plasma, their plane of polarization rotates if the magnetic field component parallel to the direction of the light propagation is nonzero, with the amount of rotation increasing with the square of the wavelength. Since the rate of rotation depends both on the density of electrons and the strength and direction of the magnetic field through which the radio waves travel, Faraday rotation can be used to investigate these quantities. Quantitavely, the amount of Faraday rotation induced can be expressed as $\Delta\chi=\lambda^2$RM, where $\lambda = c/f$ is the wavelength of the radio signal and RM is the rotation measure, given by
\begin{equation}
{\rm RM}={e^3\over{2\pi{m_e}^2c^4}}\int_{\rm LOS}{n_e}{\bf B}\cdot{\bf ds}.
\end{equation}
In the above expression, $n_e$ is the electron density, {\bf B} is the ma\-gne\-tic field, {\bf ds} is the vector incremental path length, $e$ and $m_e$ are the charge and mass of the electron, and $c$ is the speed of light. The observed RM is thus an integral along the line-of-sight (LOS) and depends critically on both the distributions of the magnetic field and electron density on the path between the observer and the polarized radio source. A comparison of the Faraday RMs of a sample of radio sources viewed through the solar corona to the RMs of the same sources observed far from the Sun reveals the presence of excess Faraday rotation that can be interpreted as originating in the solar corona.

Prior work has involved monitoring the variation of li\-ne\-ar\-ly polarized S band (2.3 GHz) radio signals as they were occulted by the solar corona from interplanetary probes such as Pioneer VI (Stelzried et al. 1970) and Helios (Hollweg et al. 1982; P{$\rm \ddot a$}tzold et al. 1987). Double-frequency decimetric observations of the Faraday rotation of polarized extragalactic radio sources have also been exploited successfully (Sakurai \& Spangler 1994; Mancuso \& Spangler 1999, 2000; Spangler \& Mancuso 2000; Spangler 2005; 
Ingleby et al. 2007) 
using the Very Large Array (VLA) radio telescope operated by the U.S. National Radio Astronomy Observatory (NRAO). This last technique has the advantage that there are always a number of po\-la\-ri\-zed radio sources around the Sun during the year that may be used for this purpose, with the possibility of sounding a {\it constellation} of different heliographic latitudes even far from the ecliptic. 
In 1997, Mancuso \& Spangler (2000, hereafter MS) analyzed the variation of the linear polarization of several extragalactic radio sources observed at decimeter (18 and 20 cm) wavelengths with the VLA radiotelescope (see Fig. 1) when they were occulted by the corona during solar minimum conditions. The main goal of that project was to establish whether a plausible global coronal model could account for the excess RMs observed at a number of simultaneous lines of sight that cut different paths through the rotating solar corona. In that work, the coronal plasma density distribution was modelled by a dense, warped equatorial component that featured an exponential decrease of density with heliographic latitude from the streamer belt and a more tenuous coronal hole component. The global coronal magnetic field distribution was instead approximated by a radially-aligned coronal magnetic field described by a functional form that included a dipolar term, a solar wind component, and a field reversal at the equator. The best model results were obtained for lines of sight at low heliographic latitudes, where the observed RMs were mostly attributed to the contribution of the steep density gradients created by the warped streamer belt. At higher heliographic latitudes, however, the predicted  RMs systematically underestimated the observed ones. MS interpreted the above anomalies as due to possible contributions of both very long wavelength Alfv\'en waves and otherwise undetected coronal structures, although unaccounted azimuthal field gradients along the line of sight produced by a tilted, non-radial, global magnetic field distribution might have been a crucial factor. 

In this work, we try to reproduce the observed RMs du\-ring solar minimum conditions at high heliographic latitudes by applying both a refined analytical global magnetic field to\-po\-lo\-gy that accounts for the possible presence of not-negligible a\-zi\-mu\-thal field gradients along the line of sight due to the combined effects of a tilted magnetic axis and a non-radial field distribution, and a more appropriate coronal hole electron density di\-stri\-bu\-tion. By fine-tuning the set of model parameters that better describe the May 1997 high-heliolatitude observations, we are able to assess the tilt of the solar magnetic axis with respect to the rotation axis. This result can be directly compared with analogous estimates obtainable from the computation of the low-order spherical harmonic expansion coefficients of the photospheric field.
\section{Model}
The global heliospheric magnetic field has the simplest form du\-ring the minimum phase of the solar cycle. If the solar magnetic field were purely dipolar, and the dipole axis were aligned with the solar rotation axis, the current sheet would simply lie in the solar equatorial plane, separating northern and southern hemispheres of opposite magnetic polarities. Not only is the current sheet tilted at {\it some} angle with respect to the solar equatorial plane but, even during solar minimum conditions, there are contributions from low-order multipole fields whose importance decreases with height.
It follows that a radially aligned, dipole-like approximation, as applied in the previous investigation by MS for the interpretation of the RM observations described above, can be inadequate to express the complexity of the coronal ma\-gne\-tic field topology and its influence on the magnitude of the observed excess RMs. 
However, since the present work is focused on the analysis of data related to lines of sight whose mi\-ni\-mum distance from Sun center is above 4 - 5 $R_{\odot}$, the effect of the tilt of the magnetic axis with respect to the rotation one is expected to be much more significant than the effect of low-order multipole components. This latter effect should be particularly important (and consequently our model may be more suitable) when analyzing Faraday rotation data in the inner and middle corona such as those collected, near solar minimum conditions, from the solar occultation of the Cassini spacecraft (Jensen et al. 2005). 

\begin{figure}
\center
\includegraphics[angle=0,width=6.8cm]{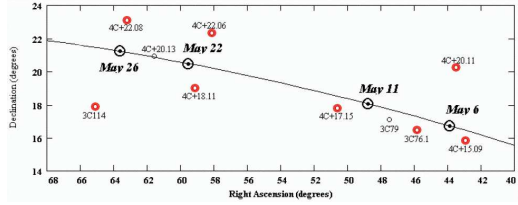}
\caption{Position of the sources (open circles) with respect to the Sun (circled points) at transit on the four days of observation. The line represents the ecliptic. Adapted from Mancuso \& Spangler (2000). Thicker circles indicate the positions of the eight sources considered in this work.}
\end{figure}
\begin{figure}
\center
\includegraphics[angle=0,width=5.8cm]{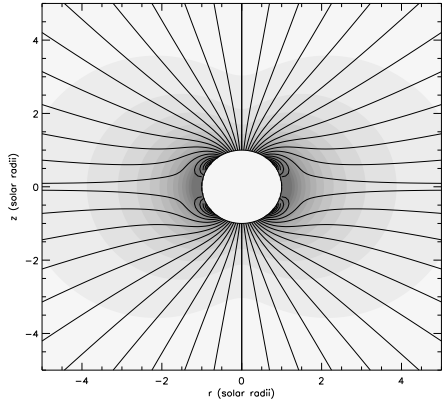} 
\caption{The magnetic field lines for the DQCS model (Banaszkiewicz et al. 1998) with the last closed field line intersecting the surface of the Sun at $60^{\circ}$ of heliolatitude. Superimposed in tonalities of gray, isodensity curves from the adopted electron density distribution.}
\end{figure}
\begin{figure}
\center
\includegraphics[angle=90, height=4.6cm]{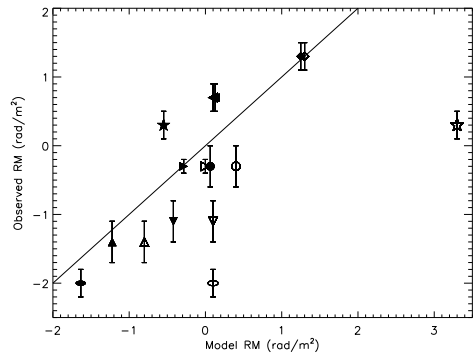} 
\caption{Observed RMs vs. model RMs for eight lines of sight at high heliolatitudes (filled symbols), compared with model results from Mancuso \& Spangler (2000) (open symbols). Different symbols refer to different lines of sight.}
\end{figure}
\subsection{Coronal magnetic field model}
In order to set up a global magnetic field topology that could adequately reproduce the observed excess RMs,   we considered the semi-empirical, cylindrically symmetric dipole-quadrupole-current-sheet (DQCS) global magnetic field model proposed by Banaszkiewicz et al. (1998), extended to account for both rotational and LOS effects related to the tilt between the solar magnetic and rotation axes. In this model, the magnetic field is expressed analytically as a superposition of a dominant dipole and a weaker quadrupole component plus a field caused by the current sheet. The contribution of the higher-order multipole components (of octupole order and above) can be  safely ignored since such fields fall off rapidly with height, thus having little influence on the global magnetic structure of the solar corona at the heights relevant to Faraday rotation observations. In cylindrical polar coordinates, $\rho$ and $z$, the DQCS model can be expressed by means of its field components as follows:
\begin{eqnarray}
\frac{B_\rho}{M} & = & \frac{3 \rho z}{r^5} + \frac{15Q}{8} \rho z \frac{4z^2-3\rho^2}{r^9} + \frac{\rho}{C(\rho,z,a_1)}\, , \\
\frac{B_z}{M}     & = & \frac{2 z^2-\rho^2}{r^5} + \frac{3Q}{8} \frac{8 z^4 + 3\rho^4-24\rho^2 z^2}{r^9} + \frac{|z|+a_1}{C(\rho,z,a_1)} \, ,
\end{eqnarray}
where $r^2 = \rho^2 + z^2$, in solar radii ($R_{\odot}$), and $C(\rho,z,a_1) \equiv a_1[(|z|+a_1)^2+\rho^2]^{3/2}$. The model has cylindrical symmetry around the dipole axis and includes three parameters ($M$, $a_1$, $Q$), two of which ($M$, $a_1$) can be constrained by normalizing the field strength to 3.1 nT at 1 AU (Forsyth et al. 1995) and as\-su\-ming the polar coronal hole boundaries at a solar latitude of $60^{\circ}$ (e.g., Axford 1976). 
The third parameter ($Q$) expresses the importance of the quadrupolar component of the field. A $Q$ value $>$ 1.5 would lead to field lines in the CS region unconnected  to the Sun, which is an unexpected feature, at least in steady state conditions. On the other hand, a $Q$ value much lower than 1.5 would not match the observational constraints (Banaszkiewicz et al. 1998; Cranmer et al. 1999), so that $Q = 1.5$ represents a rea\-so\-na\-ble choice for this parameter. The DQCS model, and the related parameter choice, are primarily intended to provide a description of polar coronal holes, and the RMs considered in the present work refer to lines of sight intersecting these high-latitude regions. In spite of its simplicity, this model has been successful in reproducing the brightness distribution of the solar corona seen by the LASCO/SoHO coronagraphs (Brueckner et al. 1995) at solar activity minimum (Schwenn et al. 1997).    
In Fig. 2, we show the field line mapping according to the DQCS model. The field is dominated by super-radial expansion of the lines rooted in polar coronal holes, whereas in the equatorial region the field lines converge above the region of closed field structures created by the dipole-quadrupole components. At large distances, above about 10 $R_{\odot}$, the field expands radially. 
The above cylindrically symmetric magnetic field model was extended to account for both rotation and LOS effects related to the tilt between the solar magnetic and rotation axes. For this purpose, eq. (2) and (3) were first expressed from cylindrical to rectangular coordinates and then transformed from the {\it magnetic} coordinate system ($x',y',z'$) (now primed for convenience) to the reference frame  ($x,y,z$) of the observer, assumed to have the same origin and two axes, $x$ and $y$, in the plane containing the Sun center, the radio source and the observer, with the $x$-axis pointing towards the observer. The above transformation can be expressed in spherical coordinates ($r,\theta,\phi$) as 
\begin{equation}
{\bf B} = {\bf T}_y(\theta) {\bf T}_z(\phi) [{\bf T}_y(\theta_B) {\bf T}_z(\phi_B)]^{-1} {\bf B'} \, , 
\end{equation}
where ${\bf T}_\xi(\psi)$ is the usual axis rotation matrix describing a rotation of an angle $\psi$ around the $\xi$ axis, $\theta=\pi/2$ along the line of sight, 
and $\theta_B$ and $\phi_B$ identify the position of the solar magnetic axis in the considered $(r,\theta,\phi)$ reference frame. Eq. (4) leads to magnetic field components $B_r$ and $B_\phi$ ($B_\theta$ does not contribute to the observed RMs) expressed as functions of $r$, of the model parameters $M$, $a_1$, $Q$,  and of the angles $\theta_B$ and $\phi_B$. 
The re\-sul\-ting formulae can be applied in a straightforward way in as far as the dipole and the rotation axes are the same. On
the other hand, if the rotation and magnetic dipole axes are tilted, further transformations have to be included. These are
needed to express $\theta_B$ and $\phi_B$ as functions of $\theta_{RB}$, the tilt of the magnetic axis with respect
to the rotation axis,
and $\phi_{RB}$, a time-varying azimuthal angle corresponding to a rigid coronal rotation at a rate $\phi_{RB} = {\phi^{\circ}_{RB}} + \Omega t$, where $\Omega$ is the Sun's angular velocity and ${\phi^{\circ}_{RB}}$ is the angle 
$\phi_{RB}$ at an arbitrary initial time, conveniently chosen as May 1st, 1997. The formulae were moreover extended to account for the effect of the Archimedean spiral (Parker 1958) whenever the solar wind becomes superAlfv\'enic. 
\begin{figure}
\center
\includegraphics[angle=90, height=4.6cm]{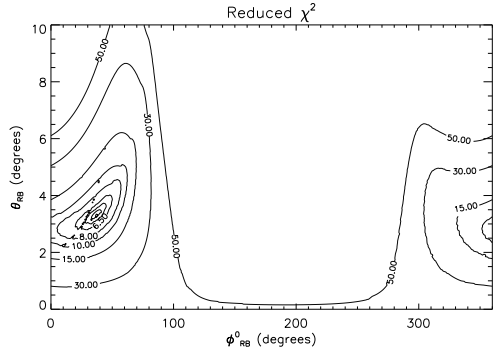} 
\caption{Contour plot of the reduced chi-squared $\chi^2_\nu$ for eight lines of sight at high heliolatitudes. $\theta_{RB}$ is the tilt of the magnetic axis with respect to the rotation axis; ${\phi^{\circ}_{RB}}$ is  the azimuthal angle that describes the orientation of the magnetic axis at an arbitrary initial time.}
\end{figure}

\subsection{Coronal electron density model}
In order to infer coronal magnetic fields from Faraday rotation measurements, 
a model for the coronal electron density distribution is required.
This information is available by means of different diagnostic tools provided by the instruments aboard the SOlar and Heliospheric Observatory (SOHO) spacecraft. At least in the coronal hole region, an appropriate electron density distribution that is in agreement with the observations (e.g., Weisberg et al. 1976; Munro \& Jackson 1977; Gallagher et al. 1999) can be described by a cosine-power law of the form:
\begin{equation}
n_e(r)  = n_H(r) +  [n_S(r) - n_H(r)]  \cos^2\theta_m \, .
\end{equation} 
Here $\theta_m$ is the angular distance of a point from the current sheet in a heliomagnetic coordinate system, $n_H(r)$  is the radial electron density distribution inside a coronal hole, and $n_S(r)$  is the radial electron density distribution inside the coronal streamer belt, both observationally constrained (see MS). 
\begin{figure}
\center
\includegraphics[angle=0,height=4.6cm]{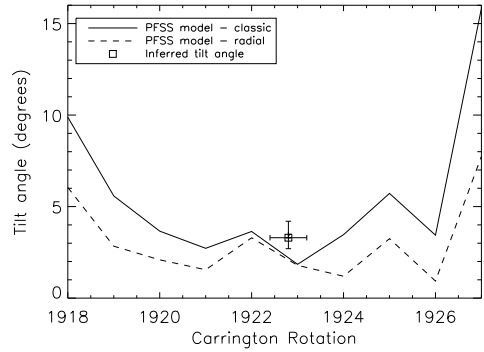} 
\caption{Tilt angle derived from the analysis of the photospheric magnetic field adopting both the classic (solid line) and the radial (dashed line) PFSS models of the WSO. The square symbol indicates the tilt angle inferred from our independent coronal rotation measurements. 
The vertical error bar corresponds to a 99\% confidence interval calculated from the $\chi^2$ surface in Fig. 4 taking into account $\theta_{RB}$ and ${\phi^{\circ}_{RB}}$ jointly.}
\end{figure}
\section{Results and conclusions}
The LOS component of the magnetic field model and the electron density distribution presented in Sect. 2 were first multiplied and then numerically integrated along the lines of sight, ac\-cor\-ding to eq. (1), in order to obtain the predicted RMs. Since this model does not allow us to take into proper account the coronal contribution to the observed excess RMs around the equatorial region (due to the presence of the steep density gradients created by the warped equatorial streamer belt), we restricted our study to lines of sight at heliolatitudes higher than $20^{\circ}$ with respect to the solar equator. With all other parameters of the model constrained, the two remaining free parameters ($\theta_{RB}$ and ${\phi^{\circ}_{RB}}$) of the model were varied within their suitable ranges ($0^{\circ} \leq \theta_{RB} \leq 30^{\circ}$ and $0^{\circ} \leq {\phi^{\circ}_{RB}} \leq 360^{\circ}$) to obtain the best agreement with the data. 
Our best fit results are shown in Fig. 3 and compared with the ones predicted by MS for the same set of radio sources. From visual inspection, it is clear that the improvement is substantial with respect to the results from the previous investigation. This is due to both the use of a more refined global magnetic field model that takes into account the tilted multipolar ma\-gne\-tic field topology of the corona and a better representation of the electron density distribution in the coronal holes. Differences between model and observed RMs are, overall, well within one rad m$^{-2}$, and might be attributed to the presence of very long wavelength Alfv\'en waves in the polar regions and/or of otherwise undetected coronal structures.

The most interesting and newer aspect of our analysis is, however, the intriguing possibility of employing coronal Faraday rotation observations to assess the tilt angle $\theta_{RB}$ of the magnetic axis with respect to the rotation axis. Our best-fitting model yields a well-defined minimum ($\chi^2_\nu = 5.7$) of the reduced chi-square between observed and predicted RMs at an inclination corresponding to $\theta_{RB}=3.3^{\circ}$ (Fig. 4). Note that the MS model, applied to the same eight sources, yielded a reduced chi-squared of about an order of magnitude larger ($\chi^2_\nu=47.8$). 
The somewhat large reduced $\chi^2$ values quoted above might have been affected both by the fact that the model did not take into account the possible presence of the already mentioned plasma irregularities or Alfv\'enic fluctuations and by the possibility of a somewhat optimistic estimate of the systematic measurement uncertainties reported by MS, which were attributed solely to radiometer noise. 
Information on the tilt of the global ma\-gne\-tic field can also be obtained by extrapolating the observed photospheric magnetic field distribution into the corona using models based upon various approximations. The most widely used is the potential field-source surface (PFSS) model (e.g., Hoeksema et al 1983; Wang and Sheeley 1992). In this model, the coronal magnetic field is assumed to be potential and is obtained by solving the Laplace equation in a spherical shell bounded by the photosphere and the source surface (usually placed at $R_{\rm ss}=2.5~R_{\odot}$), where the magnetic field is assumed to be purely radial. In Fig. 5, we show the tilt angle computed from the spherical harmonic expansion coefficients obtained by adopting two different PFSS models ({\it classic} and {\it radial}) and u\-sing Wilcox Solar Observatory (WSO) data (freely available at {\tt http://quake.stanford.edu/$\sim$wso}) along a time interval of nine Carrington Rotations centered around the period of observations under study. From Fig. 5, it is clear that the tilt angles inferred with the above technique are in fair agreement with our independent estimate obtained around CR 1923. 

In conclusion, we were able to reproduce the observed high-latitude RMs during solar minimum conditions by applying both a refined analytical global magnetic field topology that accounts for a tilted magnetic axis and a non-radial field distribution, and a more appropriate coronal hole electron density distribution. By fine-tuning the set of model parameters that better describe the observations, we were able to assess the tilt of the solar ma\-gne\-tic axis with respect to the rotation axis. This result is in sa\-ti\-sfac\-tory agreement with analogous estimates obtained from the computation of the low-order spherical harmonic expansion coefficients of the photospheric field.  Still, our result is not unambiguous, since it is dependent on the assumed, though plausible, density model and the correct choice of the magnetic field model parameters. Moreover, the assumptions of cylindrical symmetry of both the $\bf B$ field and the $n_e$ distribution, which make feasible the analytical model, place strong limitations on its use in the presence of plasma inhomogeneities, especially away from solar minimum conditions. 

\begin{acknowledgements}
WSO is supported by NASA, NSF, and ONR. NRAO is a facility of the National Science Foundation operated under cooperative agreement by Associated Universities, Inc.  SOHO is a project of international cooperation between ESA and NASA.
\end{acknowledgements}

\end{document}